\documentclass[%
 reprint,
 amsmath,amssymb,
 aps,
]{revtex4-2}

\usepackage{graphicx}
\usepackage{dcolumn}
\usepackage{bm}

\usepackage{graphicx,amsmath,amssymb,epsfig,epstopdf,color,float}
\usepackage{stfloats}
\usepackage{color}
\usepackage{lineno}
\usepackage{enumerate}
\usepackage{stfloats}
\usepackage{comment}\allowdisplaybreaks
\def\be{\begin{equation}}
\def\ee{\end{equation}}
\def\bea{\begin{eqnarray}}
\def\eea{\end{eqnarray}}
\def\bes{\begin{subequations}}
\def\ees{\end{subequations}}

\def\pl{\partial}
\def\ep{\epsilon}

\def\vep{\varepsilon}

\makeatletter
\newcommand*{\centerfloat}{%
  \parindent \z@
  \leftskip \z@ \@plus 1fil \@minus \marginparwidth
  \rightskip \leftskip
  \parfillskip \z@skip}
\makeatother

\begin{document}


\title{Quantum Reductive Perturbation Method for Photon Propagations in a Cold Atomic Gas}

\author{Yao Ou$^{1}$, and Guoxiang Huang$^{1,2}$
       }
       \homepage{gxhuang@phy.ecnu.edu.cn}
\affiliation{$^1$State Key Laboratory of Precision Spectroscopy,
                 East China Normal University, Shanghai 200062, China\\
             $^2$NYU-ECNU Joint Institute of Physics, New York University at Shanghai, Shanghai 200062, China
             }

\date{\today}

\begin{abstract}
We develop a quantum reductive perturbation method (RPM), a generalization of classical RPM widely used in nonlinear wave theory, to derive a simplified model (i.e. quantum nonlinear Schr\"odinger equation) from fully quantum Heisenberg-Langevin-Maxwell equations describing photon propagations in a coherent cold atomic gas. The result is used to discuss two-photon bound states and optical solitons in the gas.
Though a specific system is considered, the quantum RPM established here is very general and can be applied to other complex quantum nonlinear problems.
\end{abstract}

\maketitle


\section{\label{sec:level1}Introduction}
In 1960, Gardner and Morikawa~\cite{Gardner1960} introduced a scaling transformation combined with a perturbation expansion for studying asymptotic behaviors of nonlinear waves. Their perturbation technique was later developed generally by Taniuti and his colleagues, named as reductive perturbation method (RPM)~\cite{Taniuti1974}. In past decades, RPM has become a powerful and systematic way for reducing complex classical nonlinear problems to simplified models [e.g., Korteweg–de Vries equation, nonlinear Schr\"odinger equation (NLSE), etc.]~\cite{Jeffrey1982}, widely used in various systems, especially in fluid dynamics and nonlinear optics~\cite{Newell1992}.

In recent years, great efforts have been paid to the research of quantum nonlinear optics~\cite{Chang2014,Drummond2014,Quesada2022}, for which one often needs to solve fully quantized many-body problems that involve many quantum fields describing photons and atoms (or other quantum emitters) coupled strongly. Analytical approaches by using conventional perturbation methods to solve such problems are usually invalid; numerical simulations also become difficult because the dimension of Hilbert state space grows
exponentially as the particle number of the system increases. Thus it is desirable to establish a quantum RPM, which can reduce complex quantum many-body problems into simplified models that are easy to solve analytically and/or numerically. However, there is no such a method reported up to now.

In this Letter, we develop a quantum RPM  by taking a cold three-level atomic gas coupled with laser fields as an example. By generalizing the classical RPM~\cite{Jeffrey1982,Newell1992}, we derive a quantum NLSE from fully-quantized Heisenberg-Langevin-Maxwell equations (which have seven quantum fields coupled each other) governing the quantum dynamics of the probe field envelope and the atoms under the condition of electromagnetically induced transparency (EIT)~\cite{Fleischhauer2005}. The result obtained is used to discuss the two-photon bound states and slow-light solitons in the system.

Before preceding, we notice that there exist some techniques for deriving quantum NLSE in nonlinear optical systems, including optical fiber~\cite{Drummond1987,Lai1989}, alkali vapor in a high-Q Fabry-P\'erot cavity~\cite{Deutsch1992}, Kerr liquid in a waveguide~\cite{Deutsch1993}, dispersive dielectric~\cite{Larre2015}, ladder-type atomic gas~\cite{Gullans2016}, and lambda-type (EIT) atomic gas, etc.~\cite{Zhu2021}.  In these derivations, one either derived a classical NLSE from coupled
classical nonlinear equations firstly and quantized the classical NLSE finally~\cite{Drummond1987,Lai1989,Deutsch1992,Deutsch1993,Larre2015,Gullans2016}, or employed a phenomenological approach to construct quantum NLSE through separate estimations of dispersion and nonlinearity~\cite{Zhu2021}. The derivation developed in the present work is very different from them, because the original coupled field equations are fully quantized at the beginning of the derivation and they are reduced into a single quantum NLSE in a standard and systematic way, well in line with the spirit of RPM~\cite{Gardner1960,Taniuti1974,Jeffrey1982,Newell1992}.

Moreover, in our approach the quantized optical field envelope obeys equal-time commutation relations. This is more suitable from the viewpoint of standard canonical quantization, at variance with that obtained in previous studies~\cite{Drummond1987,Lai1989,Deutsch1992,Deutsch1993,Larre2015,Gullans2016,Zhu2021} where commutation relations are equal-space ones. In addition, based on the quantum RPM the dynamical evolution of the atomic transition operators can also be obtained simultaneously when the solutions of the quantum NLSE are gained. We stress that, although a particular system is chosen in our calculation, the quantum RPM established here is very general and can be generalized to other complex quantum many-body systems.

\section{Model}
\begin{figure*}[ht]
\centering
\includegraphics[width=1.5\columnwidth]{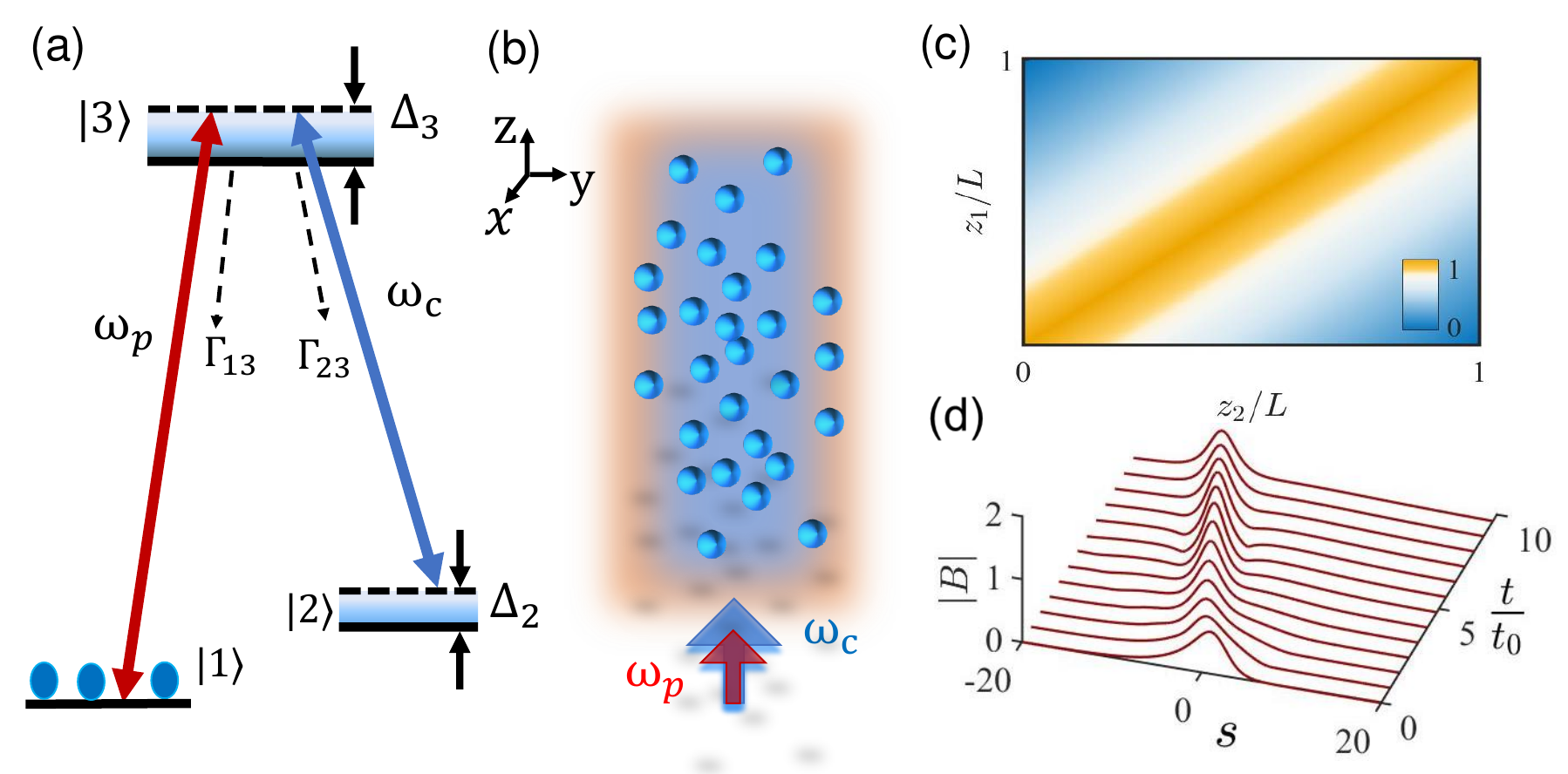}
\caption{(a)~Energy-level diagram and excitation scheme of the EIT-based $\Lambda$-shaped three-level atomic gas.
For the explanation of the quantities indicated in the figure, see text.
(b)~Possible experimental geometry. To reduce Doppler effect, both the probe and control fields are assumed to propagate along $z$ direction.
(c)~Probability distribution $|\Phi|^2$ of two-photon bound state in phase space as a function of $z_1/L$ and $z_2/L$.
(d)~Optical soliton in real space as a function of
$s=(z-V_g t)/l_0$ and $t/t_0$.
}
\label{Fig1}
\end{figure*}

We consider a cold atomic gas with $\Lambda$-shaped three-level configuration, with two ground states $|1\rangle$ and $|2\rangle$ and an excited state $|3\rangle$; see Fig.~\ref{Fig1}(a).

A pulsed probe laser field (frequency $\omega_p$, wavenumber $k_p$) couples the transition $|1\rangle\leftrightarrow|3\rangle$; a strong continuous-wave control laser field (frequency $\omega_c$, wavenumber $k_c$) couples  the transition $|2\rangle\leftrightarrow|3\rangle$.
$\Delta_{2}=\omega_{p}-\omega_{c}-(\omega_{2}-\omega_{1})$ and $\Delta_{3}=\omega_{p}-(\omega_{3}-\omega_{1})$ are respectively two- and one-photon detunings, with $\hbar\omega_{\alpha}$ the eigenenergy of the state $|\alpha\rangle$; $\Gamma_{\alpha3}$ is the decay rate of the spontaneous emission from $|3\rangle$ to $|\alpha\rangle$ ($\alpha=1,2$). Solid blue dots in the figure mean that the atoms are initially prepared at the ground state $|1\rangle$.

The total electric field of the system is given by
${\hat{\bf E}}={\hat{\bf E}}_{p}+{\bf E}_{c}$.
Here, ${\hat{\bf E}}_{p}\equiv {\bf e}_{p}{\cal E}_{p}{\hat E}_{p}({\bf r},t)e^{i({\bf k}_{p}\cdot{\bf r}-\omega_{p}t)}$+h.c. is the quantized probe field, with ${\bf e}_{p}$ the  unit polarization vector, ${\cal E}_{p}\equiv[\hbar\omega_{p}/(2\varepsilon_{0}V)]^{1/2}$ the field amplitude of single photon, and $V$ the quantization volume; ${\bf E}_{c}\equiv {\bf e}_{c}{\cal E}_{c}\,e^{i({\bf k}_{c}\cdot{\bf r}-\omega_{c}t)}$+c.c. is the control field with unit polarization vector ${\bf e}_{c}$ and field amplitude ${\cal E}_{c}$,
assumed to be strong enough and can be taken as a classical and undepleted one.
h.c. (c.c.) denotes Hermitian (complex) conjugate.
The annihilation operator of probe photons, $\hat{E}_p({\bf r},t)$, is a slowly-varying function of ${\bf r}$, $t$ and  obeys the equal-time commutation relation
\be \label{CR0}
\left[\hat{E}_p({\bf r},t), \hat{E}_p^{\dag}({\bf r}',t)\right]=V \delta ({\bf r}-{\bf r}').
\ee
In order to suppress Doppler effect, we assume both the probe and control fields propagate along $z$ direction [see Fig.~{\ref{Fig1}(b)}].

Under electric-dipole, rotating-wave, and paraxial approximations, the Hamiltonian of the system reads
\begin{align}\label{Hamiltonian}
{\hat H}&=\int d^3r\bigg[-\frac{\hbar c}{V}\hat{E}_{p}^{\dag}\left(i\frac{\partial}{\partial z}\right)\hat{E}_{p} -\frac{\hbar c}{2k_pV}\hat{E}_{p}^{\dag}\nabla_{\perp}^2 \hat{E}_{p}
\notag \\
&\hspace{4mm}-\frac{\hbar N}{V}\left(\sum_{\alpha=2,3}\Delta_{\alpha}\hat{S}_{\alpha\alpha}+g_{p}\hat{S}_{31}^\dag\hat{E}_{p}+\Omega_c\hat{S}_{32}^\dag+{\rm h.c.}\right)\bigg],
\end{align}
where $ d^3r=dx dy dz$, $N$ is the total atomic number,
$\hat{S}_{\alpha\beta}({\bf r},t)=|\beta\rangle \langle \alpha|
\exp\{i[(k_\beta-k_\alpha)\cdot{\bf r}-(\omega_\beta-\omega_\alpha
+\Delta_\beta-\Delta_\alpha)t]\}$
is the atomic transition operator from the state $|\alpha\rangle$ to $|\beta\rangle$, satisfying $\left[\hat{S}_{\alpha\beta}({\bf r},t),\hat{S}_{\mu\nu}({\bf r}',t) \right]=(V/N)\delta({\bf r}-{\bf r}')[\delta_{\alpha\nu}\hat{S}_{\mu \beta}({\bf r},t)-\delta_{\mu \beta}\hat{S}_{\alpha\nu}({\bf r},t)]$. Here, $k_1=0$, $k_2=k_p-k_c$, and $k_3=k_p$; $\Omega_c=(\mathbf{e}_c\cdot \mathbf{p}_{32})\mathcal{E}_c/\hbar$ is the half Rabi frequency of the control field; $g_{p}=(\mathbf{e}_p\cdot\mathbf{p}_{31})\mathcal{E}_p/\hbar$ is the coefficient denoting the strength of the coupling between the probe photon and the transition $|1\rangle\leftrightarrow|3\rangle$;  $\mathbf{p}_{\alpha\beta}$ is the electric dipole matrix element associated with the transition from $|\beta\rangle$ to $|\alpha\rangle$.

The dynamics of the system is governed by the Maxwell and Heisenberg-Langevin (MHL) equations
\begin{subequations}\label{MHL}
\begin{align}
& i\left(\frac{\partial}{\partial z}+\frac{1}{c}\frac{\partial}{\partial t}\right)\hat{E}_{p}+\frac{1}{2k_p}\nabla_{\perp}^2 \hat{E}_{p}+\frac{g_{p}^\ast N}{c}{\hat S}_{31}=0, \label{MHL1}\\
& i\frac{\partial}{\partial t}{\hat S}_{\alpha\beta}=\left[{\hat S}_{\alpha\beta},\frac{{\hat H}}{\hbar}\right]-i{\hat{\cal L}}({\hat S}_{\alpha\beta})+i{\hat F}_{\alpha\beta}, \label{MHL2}
\end{align}
\end{subequations}
where ${\hat{\cal L}}({\hat{S}_{\alpha\beta}})$ is the $3\times3$ relaxation matrix, ${\hat F}_{\alpha\beta}$ are $\delta$-correlated Langevin noise operators.
Explicit expressions of the MHL equations are presented in Sec.~S1\,B of {\color{blue}Supplement 1}. Note that these equations are fully quantized and contain seven independent quantum fields (i.e., ${\hat E}_{p}$,  ${\hat S}_{11}$, ${\hat S}_{22}$, ${\hat S}_{33}$, ${\hat S}_{21}$, ${\hat S}_{31}$, ${\hat S}_{32}$) that are coupled nonlinearly.

The model described above can be realized by many atomic ensembles. One of candidates is the laser-cooled alkali $^{87}$Rb gas, with levels chosen to be $|1\rangle=|5^2S_{1/2},F=1,m_{F}=1\rangle$, $|2\rangle=|5^2S_{1/2},F=2,m_{F}=1\rangle$ and $|3\rangle=|5^2P_{3/2},F=2,m_{F}=1\rangle$. System parameters (used in the following calculations) are given by $\Gamma_{13}=\Gamma_{23}=2\pi\times3\,{\rm MHz}$, $\Delta_3=2\pi\times60\,{\rm MHz}$, $\Delta_2=2\pi\times1.2\,{\rm MHz}$, ${\cal N}_{a}\equiv N/V=8\times10^{10}\,{\rm cm}^{-3}$, and $\Omega_c=2\pi\times28$MHz.

\section{Quantum NLS equation and commutation relations}
In order to reduce Eqs.~(\ref{MHL1}) and (\ref{MHL2}), we make the asymptotic expansion $\hat{E}_p=\sum_{j=1}^{\infty} \varepsilon^j \hat{E}_p^{(j)}$,
$\hat{S}_{\alpha \beta}=\sum_{j=0}^{\infty} \varepsilon^j \hat{S}_{\alpha \beta}^{(j)}$, $\hat{F}_{\alpha \beta}=\sum_{j=3}^{\infty} \varepsilon^j \hat{F}_{\alpha \beta}^{(j)}$.
To obtain a divergence-free expansion, all quantities on the right hand side of the expansion are taken as functions of the multi-scale variables $z_l=\varepsilon^l z$  ($l=0,1,2$), $t_l=\varepsilon^l t$ ($l=0,1$), $x_1=\varepsilon x$, and $y_1=\varepsilon y$~\cite{Newell1992}, where $\varepsilon$ is a small parameter characterizing the  probe-field amplitude.
Substituting such an expansion into the MHL equations (\ref{MHL}) and comparing the powers of $\varepsilon$, we obtain a series of linear but inhomogeneous operator equations for
$\hat{E}_p^{(j)}$ and $\hat{S}_{\alpha \beta}^{(j)}$ that can solved order by order. Detailed expressions of these equations are given in Sec.~S2\,A of {\color{blue}Supplement 1}.

The expansion equations can be divided into two groups. The group-I equations, which are for  $\hat{E}_p^{(j)}$, $\hat{S}_{21}^{(j)}$, $\hat{S}_{31}^{(j)}$ and have non-zero solutions starting from the first order, read

\begin{subequations}\label{RHLM}
\begin{align}
i &\left(\frac{\partial}{\partial z_0}+\frac{1}{c} \frac{\partial}{\partial t_0}\right)\hat{E}_p^{(j)}+\frac{g_p^{*}N}{c} \hat{S}_{31}^{(j)}=\hat{M}^{(j)}, \\
&\left(i \frac{\partial}{\partial t_0}+d_{21}\right) \hat{S}_{21}^{(j)}+\Omega_c^* \hat{S}_{31}^{(j)}=\hat{N}^{(j)},  \\
&\left(i \frac{\partial}{\partial t_0}+d_{31}\right) \hat{S}_{31}^{(j)}+\Omega_c \hat{S}_{21}^{(j)}+g_p \hat{E}_p^{(j)}=\hat{O}^{(j)},
\end{align}
\end{subequations}
$j=1,2,3,...$, which can be recast into the form
\begin{equation}\label{LS0}
\hat{L}\, \hat{E}_p^{(j)}=\hat{S}^{(j)},
\end{equation}
where
$\hat{L}= i\left(\frac{\partial}{\partial z_0}+\frac{1}{c} \frac{\partial}{\partial t_0}\right) \\
\left[\left|\Omega_c\right|^2-\left(i \frac{\partial}{\partial t_0}+d_{21}\right)
\left(i \frac{\partial}{\partial t_0}+d_{31}\right)\right] +\frac{g_p^{2}N}{c}\left(i \frac{\partial}{\partial t_0}+d_{21}\right)$ is a linear
operator. Explicit expressions of $\hat{M}^{(j)}$, $\hat{N}^{(j)}$, $\hat{O}^{(j)}$, and $\hat{S}^{(j)}$ are given in Sec.~S2\,A of {\color{blue}Supplement 1}.

The group-II equations, which are for $\hat{S}_{32}^{(j)}$, $\hat{S}_{11}^{(j)}$, $\hat{S}_{22}^{(j)}$, $\hat{S}_{33}^{(j)}$ and have non-zero starting from the second order ($j=2,3,...$). Their explicit expressions are also given in Sec.~S2\,A of {\color{blue}Supplement 1}. We can first solve the group-I equations to get the solutions of $\hat{E}_p^{(j)}$, $\hat{S}_{21}^{(j)}$, and $\hat{S}_{31}^{(j)}$, and then get the solutions of $\hat{S}_{32}^{(j)}$, $\hat{S}_{11}^{(j)}$, $\hat{S}_{22}^{(j)}$ and $\hat{S}_{33}^{(j)}$  by solving the group-II equations, order by order.

At the first-order approximation ($j=1$), Eq.~(\ref{LS0}) admits the solution
\begin{equation}\label{1st-solution}
  \hat{E}_{p}^{(1)}=\hat{A} e^{i\theta},  \quad  \theta=K(\omega)z_0-\omega t_0,
\end{equation}
where $\hat{A}$ is an envelope function of slow variables $x_1$, $y_1$, $z_1$, $t_1$, and $z_2$; $\omega$ is the sideband frequency of the probe pulse;
\begin{align} \label{LinDis}
K(\omega)=\frac{\omega}{c}+\frac{|g_{p}|^{2}N}{c}\frac{\omega+d_{21}}{D(\omega)}
\end{align}
is the linear dispersion relation, with $D(\omega)=|\Omega_{c}|^2-(\omega+d_{21})(\omega+d_{31})$. The atomic variables can be obtained through solving (\ref{RHLM}), given by
\begin{subequations}\label{S(1)}
\begin{align}
{\hat S}_{21}^{(1)}&=-\frac{g_p \Omega_c^*}{|\Omega_{c}|^2-d_{21} d_{31}} \hat{A} e^{i\theta}; \\
{\hat S}_{31}^{(1)}&=\frac{g_p d_{21}}{|\Omega_{c}|^2-d_{21}d_{31}} \hat{A} e^{i\theta},
\end{align}
\end{subequations}
with other $\hat{S}^{(1)}_{\alpha\beta}$ zero. Under the EIT condition $|\Omega_c|^2\gg \gamma_{21}\gamma_{31}$ a transparency window opens in the absorption spectrum  Im[$K(\omega)$] near $\omega=0$~\cite{Zhu2021}.

With the first-order solution given above, we can calculate $\hat{S}^{(2)}$. Then at the second-order approximation ($j=2$), Eq.~(\ref{LS0}) has the form $\hat{L}\, \hat{E}_p^{(2)}=-i D[\partial \hat{A}/\partial z_1+(1/V_g) \partial \hat{A}/\partial t_1] e^{i \theta}$, with $V_g\equiv [(\partial K/\partial\omega)|_{\omega=0}]^{-1}$ the group velocity of the probe pulse.
Because $e^{i \theta}$ is the eigenmode of $\hat{L}$, the term on the right hand side is a secular one that should be eliminated. Thus we take
\be \label{SC1}
i\left(\frac{\pl}{\pl z_1}+\frac{1}{V_g}\frac{\pl}{\pl t_1}\right)\hat{A}=0,
\ee
which is also called solvability condition~\cite{Jeffrey1982,Newell1992}.
Solutions of $\hat{E}_p^{(2)}$ and $\hat{S}^{(2)}_{\alpha\beta}$  are presented in
Sec.~S2\,B of {\color{blue}Supplement 1}.

With the first- and second-order solutions in hand, we can go to the third-order approximation ($j=3$). The condition of eliminating the secular term at this order
requires
\be \label{SC2}
i\frac{\pl \hat{A}}{\pl z_2}-\frac{K_2}{2}\frac{\pl^2 \hat{A}}{\pl t_1^2}+
\frac{1}{2k_p}\left(\frac{\pl^2}{\pl x_1^2}+\frac{\pl^2}{\pl y_1^2}\right) \hat{A} +W\hat{A}^{\dag}\hat{A}\hat{A}+i\hat{G}_{p}=0.
\ee
Here $K_2(\equiv (\partial^2K/\partial\omega^2)|_{\omega=0})$, $W$, $\hat{G}_{p}$ are coefficients of group-velocity dispersion,
Kerr nonlinearity, and noise operator, respectively. Their explicit expressions are also
presented in Sec.~S2\,B of {\color{blue}Supplement 1}.

To get a unified equation for $\hat{A}$, following Ref.~\cite{Newell1992} we multiply (\ref{SC2}) by $\vep$ and add it to (\ref{SC1}), which yields

\bea \label{NLSE100}
&& i\left(\frac{\pl}{\pl Z}+\frac{1}{V_g}\frac{\pl}{\pl t_1}\right)\hat{A}
+\vep \left[-\frac{K_2}{2}\frac{\pl^2 }{\pl t_1^2}\hat{A}+ \right. \nonumber\\
&& \left.\frac{1}{2k_p}\left(\frac{\pl^2}{\pl x_1^2}+\frac{\pl^2}{\pl y_1^2}\right)\hat{A}+W\hat{A}^{\dag}\hat{A}\hat{A}+i\hat{G}_{p}\right]=0.
\eea
where $\pl/\pl Z\equiv \pl/\pl z_1+\vep \pl/\pl z_2$~\cite{Newell1992}.
Taking $\hat{B}=\ep \hat{A}$, returning the original variables, and then assuming $\hat{B}=\hat{B}(x,y,\xi,t)$  (with $\xi=z-V_g t$), Eq.~(\ref{NLSE100}) becomes
$i\frac{1}{V_g}\frac{\pl}{\pl t}\hat{B}
-\frac{K_2}{2}\left(V_g^2 \frac{\pl^2 }{\pl \xi^2}-2V_g \frac{\pl^2 }{\pl \xi\pl t}+\frac{\pl^2 }{\pl t^2}\right)\hat{B}
+\frac{1}{2k_p}\left(\frac{\pl^2}{\pl x^2}+\frac{\pl^2}{\pl y^2}\right)\hat{B}
+W\hat{B}^{\dag}\hat{B}\hat{B}+i\hat{G}_{p}=0$.
Because $\hat{B}\approx \ep$, $\pl /\pl t \approx \ep^2$, $\pl /\pl \xi \approx \ep$, $\pl /\pl x \approx \ep$, $\pl /\pl y \approx \ep$, the terms involve $\pl^2/\pl \xi\pl t$ and $\pl^2/\pl t^2$ are high-order ones and hence can be neglected. Thus we have
\bea \label{QNLSE5}
&& i\frac{1}{V_g}\frac{\pl}{\pl t}\hat{B}
-\frac{K_2}{2}V_g^2 \frac{\pl^2 }{\pl \xi^2}\hat{B}
+\frac{1}{2k_p}\left(\frac{\pl^2}{\pl x^2}+\frac{\pl^2}{\pl y^2}\right)\hat{B}\nonumber\\
&& +W\hat{B}^{\dag}\hat{B}\hat{B}+i\hat{G}_{p}=0.
\eea
It is a (3+1)D quantum NLSE with time $t$ as an evolution variable. Since under the EIT condition $\hat{G}_{p}$ is very small and the imaginary parts of $K_2$, $V_g$, and $W$ are much smaller than their real parts~\cite{Zhu2021},  we shall disregard $\hat{G}_{p}$ and the imaginary parts of $K_2$, $V_g$, and $W$ in the following.

Using the solution (\ref{1st-solution}), we have
$[\hat{E}_{p}(\mathbf{r},t),\hat{E}_{p}^{\dag}(\mathbf{r}',t)]
\approx [\vep\hat{A}({\bf r},t) e^{i\theta(z,t)},\vep\hat{A}^{\dag}({\bf r}',t) e^{-i\theta(z',t)}]
=[\hat{B}({\bf r},t), \hat{B}^{\dag}({\bf r}',t)]e^{iK(z-z')}$. From the commutation relation of $\hat{E}_p$ given by
(\ref{CR0}) we see that $\hat{B}$ satisfies the equal-time commutation relation
\be
[\hat{B}({\bf r},t), \hat{B}^{\dag}({\bf r}',t)]=V\delta(\mathbf{r}-\mathbf{r}'),
\ee
with ${\bf r}=(x,y,\xi)$. It is different from the equal-space commutation relation used in Refs.~\cite{Drummond1987,Lai1989,Deutsch1992,Deutsch1993,Larre2015,Gullans2016,Zhu2021}.
Since in (1+1)D case the commutation relation of $\hat{E}_{p}$ is given by $[\hat{E}_{p}(z,t),\hat{E}_{p}^{\dag}(z',t)] =L\delta(z-z')$ ($L$ is the length of the system), the commutation relation for $\hat{B}$ is thus given by $[\hat{B}(\xi, t), \hat{B}^{\dag}(\xi',t)]=L\delta(\xi-\xi')=L\delta(z-z')$. For a detailed discussion on these commutation relations, see Sec.~S2\,D of {\color{blue}Supplement 1}.

\section{Two-photon bound states}
In principle, the quantum NLSE (\ref{QNLSE5}) can be used to study photonic excitations in the system with arbitrary number of photons. Here we discuss only two cases, i.e. the case of two photons and the case of a large number of photons.
For simplicity, we assume the atomic gas is filled in a cigar-shaped cell or trapped in a photonic waveguide~\cite{Chang2014}, and hence the diffraction term $(1/2k_p)(\pl^2/\pl x^2+\pl^2/\pl y^2)\hat{B}$ can be neglected, thus we have
\be \label{QNLSE6}
i\frac{1}{V_g}\frac{\pl}{\pl t}\hat{B}
-\frac{K_2}{2}V_g^2 \frac{\pl^2 }{\pl \xi^2}\hat{B}+W\hat{B}^{\dag}\hat{B}\hat{B}=0.
\ee

For the first case, we are interested in two-photon bound states.
In Schr\"{o}dinger picture, the system described by the (1+1)D quantum NLSE (\ref{QNLSE6}) is controlled by the Schr\"{o}dinger equation
\be\label{SchroEq}
i \hbar \frac{\partial}{\partial t}|\Psi(t)\rangle=\hat{H}_s|\Psi(t)\rangle,
\ee
where the effective Hamiltonian is given $\hat{H}_s=-\hbar \int d\xi\, \hat{h}_s$, with $\hat{h}_s=(K_2 V_g^3/2L)\partial \hat{B}^{\dagger}/\partial \xi \,\partial \hat{B}/\partial \xi +(V_g W/2L)\hat{B}^{\dagger} \hat{B}^{\dagger} \hat{B} \hat{B}$.
The state vector for two photons in the system has the form
$|\Psi(t)\rangle=\frac{1}{2} \int d \xi_1 d \xi_2 \Phi \left(\xi_1, \xi_2, t\right) \hat{B}^{\dagger}\left(\xi_1\right) \hat{B}^{\dagger}\left(\xi_2\right)\,|0\rangle$, with $|0\rangle$ the vacuum state and
$\int\left|\Phi \left(\xi_1, \xi_2, t\right)\right|^2 d \xi_1 d \xi_2=1$. It is easy to show that the two-photon wavefunction $\Phi$ satisfies
\be \label{SchrodET}
i \frac{\partial}{\partial t} \Phi =\left[\frac{K_2 V_g^3}{2L} \left(\frac{\partial^2}{\partial \xi_1^2}+\frac{\partial^2}{\partial \xi_2^2}\right)- \frac{V_g W}{L} \delta\left(\xi_2-\xi_1\right)\right] \Phi.
\ee
Assuming $\Phi=\phi (\xi_1,\xi_2) e^{-i E_T t}$,
we get the eigenequation
$[(K_2 V_g^3)/2L) (\partial^2/\pl  \xi_1^2+\partial^2/\partial \xi_2^2)-(V_g W/L) \delta(\xi_2-\xi_1)] \phi
=E_T \phi$,
with $E_T$ the eigenenergy. After finding the bound state solution of this equation, we obtain
\begin{align} \label{BoundSt}
&\Phi \left(z_1, z_2,t\right)=\sqrt{\zeta_0} \mathrm{e}^{-\zeta_0 |z_1-z_2|} \mathrm{e}^{-i p_0 (z_1+z_2)/2} e^{-i(E_T-p_0 V_g)t},
\end{align}
where $\zeta_0=-m_0a_0/2$, $m_0=-L/(K_2 V_g^3)$, $a_0=-V_g W/L$, $p_0$ is an arbitrary real constant denoting the momentum of the center of mass (COM). The total energy of the two-photon bound state is given by
$E_T=p_0^2/(4m_0)-m_0a_0^2/4$, with the first (second) term the energy of COM motion (two-photon bounding). A detailed discussion on two-phootn bound-states is given in Sec.~S3 of {\color{blue}Supplement 1}.

Note that the bound-state solution given above requires $m_0<0$ and $a_0>0$ (i.e. $K_2>0$, $W<0$). This condition (under which the effective interaction between the two photons is attractive) can be satisfied in the system. Based on the system parameters given previously, together with the linear dispersion relation (\ref{LinDis}),
we obtain $K_2\approx 4.82\times10^{-15}\,{\rm cm}^{-1} {\rm s}^2$ and $W\approx-2.28\times10^{-7}\,{\rm cm}^{-1} $, and hence
we have $m_0\approx -1.08\times10^{-6}\,{\rm cm}^{-1} {\rm s}$ and $a_0\approx 1.31\,{\rm cm}^{-1} {\rm s}^{-1}$.  Thereby the two-photon bound state is indeed possible in the system. Figure~\ref{Fig1}(c) shows the probability distribution $|\Phi|^2$ of the two-photon bound state as a function of $z_1/L$ and $z_2/L$.

\section{Optical solitons in semi-classical limit}
We now turn to discuss the second case, i.e. there are a large number of photons in the system. In this regime, solitons are possible, and in fact there have been many efforts paid to the study on classical solitons in EIT systems (see Ref.~\cite{Bai2019} and references therein). Such solitons can be obtained in the present theoretical scheme by taking a semi-classical limit, i.e.  by assuming the probe field is in a coherent state $|\Psi_c\rangle$. In this case the (1+1)D quantum NLSE (\ref{QNLSE6}) becomes a classical NLS equation
\be \label{QNLSE7}
i\frac{1}{V_g}\frac{\pl}{\pl t}{B}-\frac{K_2}{2}V_g^2 \frac{\pl^2 }{\pl \xi^2}B+W|B|^2B=0,
\ee
with $B=\langle \Psi_c|\hat{B}|\Psi_c\rangle$.

We are interested in bright soliton solutions in the system, which requires $K_2>0$ and $W<0$.  The single soliton solution of Eq.~(\ref{QNLSE7}) reads
\bea \label{SolitonSolu}
&& B=2 \eta_0 B_0  {\rm sech}
\left\{\frac{2\eta_0}{l_0} \left[z-\left(V_g+\frac{4\xi_0 l_0}{t_0}\right) t-z_0 \right]
\right\} \,e^{i\Theta_B},\nonumber\\
&& \Theta_B=-\frac{2\xi_0}{l_0} z +2\left[\frac{\xi_0V_g}{l_0}+2\frac{\xi_0^2-\eta_0^2}{t_0} \right] t-\phi_0.
\eea
Here $B_0=[-2/(WV_g t_0)]^{1/2}$, $l_0=[K_2V_g^3 t_0/2]^{1/2}$, $t_0\equiv 2\eta_0 /l_0$ is typical pulse duration; $\eta_0$, $\xi_0$, $s_0$, and $\phi_0$ are free real parameters.

With the soliton solution (\ref{SolitonSolu}), Up to the first-order approximation the probe field has the form
\be
{\bf E}_{p}={\bf e}_{p}\, 2E_{p0}\, {\rm sech}[\Xi (z,t)]\,\cos \Theta_s (z,t),
\ee
with $\Xi\equiv (2\eta_0/l_0)[z-(V_g+4\xi_0 l_0/t_0) t-z_0]$,
$\Theta_s\equiv(k_p+K_0-2\xi_0/l_0)z-[\omega_p  -2(\xi_0V_g/l_0+2(\xi_0^2-\eta_0^2)/t_0)]t -\phi_0$,
$E_{p0}=2{\cal E}_p B_0\eta_0$.
We see that the soliton propagates with velocity $V_s=V_g +4\xi_0l_0/t_0$.
By using the system parameters given above and taking $\xi_0=0.1$ and  $t_0=2.4\times 10^{-7}$\,s, we obtain $V_s\approx 2.11\times 10^{-4} c$, much smaller than
$c$. The ultraslow propagation of the soliton is contributed from the EIT effect. Fig.~\ref{Fig1}(d) shows the optical soliton in real space as a function of  $s=(z-V_g t)/l_0$ and $t/t_0$.

With the above result, one can obtain the propagation behavior of the atomic variables.
Assuming that the initial state of the system is
$|\Psi\rangle=|1_1,...,1_N\rangle\otimes |\Psi_c\rangle$, where $|1_1,...,1_N\rangle$
the initial quantum state of the atoms (i.e. all atoms are in the state $|1\rangle$), in the case of the optical soliton excitation the average of the atomic transition operators $\hat{S}_{31}$ and $\hat{S}_{21}$ are
given by
\bes
\bea
&& \langle \hat{S}_{31}\rangle \approx  \frac{g_p d_{21} E_{p0}}{|\Omega_{c}|^2-d_{21}d_{31}} {\rm sech}[\Xi (z,t)] \,e^{i\Theta_s},\\
&& \langle\hat{S}_{21}\rangle\approx -\frac{ g_p \Omega_c^* E_{p0}}{|\Omega_{c}|^2-d_{21} d_{31}} {\rm sech}[\Xi (z,t)] \,e^{i\Theta_s},
\eea
\ees
where
$\langle \hat{O}\rangle=\langle \Psi|\hat{O}|\Psi\rangle$, with $|\Psi\rangle=|1_1,...,1_N\rangle\otimes |\Psi_c\rangle$.
Therefore, the atomic excitations have also solitonlike behavior.

\section{Summary}

We have developed a quantum RPM for reducing fully quantum MHL equations into
a simple model, i.e. quantum NLSE, and based on which we have discussed two-photon bound states and optical solitons. We have shown that by using the quantum RPM the dynamical evolution of atomic variables can also be obtained simultaneously. We stress that, though a specific system is considered here, the quantum RPM established in the present work is very general and can be applied to other complex quantum many-body problems, including the photon propagation in waveguide quantum electrodynamics.

\begin{acknowledgments}

The project was funded by the National Natural Science Foundation of China (11975098). Data underlying the results presented in this Letter are herein and/or in {\color{blue}Supplement 1}, which can be obtained from the author.

\end{acknowledgments}


\end{document}